\documentclass[conference]{IEEEtran}
\IEEEoverridecommandlockouts
\usepackage{cite}
\usepackage{amsmath,amssymb,amsfonts}
\usepackage{algorithmic}
\usepackage{graphicx}
\usepackage{textcomp}
\usepackage{xcolor}
\usepackage{comment}
\usepackage{subcaption}
\usepackage{setspace}
\usepackage{hyperref}
\def\BibTeX{{\rm B\kern-.05em{\sc i\kern-.025em b}\kern-.08em T\kern-.1667em\lower.7ex\hbox{E}\kern-.125emX}}

\begin{document}
\title{An Automatic Speech Recognition System for Bengali Language based on Wav2Vec2 and Transfer Learning\\

\thanks
}

\author{\IEEEauthorblockN{Tushar Talukder Showrav}
\IEEEauthorblockA{\textit{Dept. of Electrical \& Electronic Engineering} \\
\textit{Bangladesh University of Engineering \& Technology}\\
Dhaka, Bangladesh \\
tushartalukder9797@gmail.com}
}

\maketitle

\begin{abstract}
An independent, automated method of decoding and transcribing oral speech is known as automatic speech recognition (ASR). A typical ASR system extracts feature from audio recordings or streams and run one or more algorithms to map the features to corresponding texts. Numerous of research has been done in the field of speech signal processing in recent years. When given adequate resources, both conventional ASR and emerging end-to-end (E2E) speech recognition have produced promising results. However, for low-resource languages like Bengali, the current state of ASR lags behind, although the low resource state does not reflect upon the fact that this language is spoken by over 500 million people all over the world. Despite its popularity, there aren't many diverse open-source datasets available, which makes it difficult to conduct research on Bengali speech recognition systems. This paper is a part of the competition named ‘BUET CSE Fest DL Sprint’. The purpose of this paper is to improve the speech recognition performance of the Bengali language by adopting speech recognition technology on the E2E structure based on the transfer learning framework. The proposed method effectively models the Bengali language and achieves 3.819 score in ‘Levenshtein Mean Distance’ on the test dataset of 7747 samples, when only 1000 samples of train dataset were used to train.
\end{abstract}


\section{Introduction}
Since speech is a method of interpersonal communication, researchers have grown more and more interested in automatic speech recognition (ASR) in recent decades. ASR started out with basic systems that only responded to a handful of sounds, but it has now developed into sophisticated systems that can speak or understand in natural language. Due to the need to automate straightforward tasks that demand for human-machine interaction, ASR technology is becoming more popular. ASR is the method of generating a word sequence from speech transcription in which the speech wave's shape is the primary consideration. Due to the variety of speech signals, speech recognition is challenging. ASR is being used extensively in a variety of tasks, including stock quotes, automatic call handling, weather reporting, and enquiry systems. 

Even though there has been a ton of research in the domain of ASR in recent years, surprisingly little has been done on Bengali, despite the fact that it is one of the most widely spoken languages in the world. Despite intensive theoretical and computational research on modeling Bengali phonology and the development of powerful deep learning networks, language technologies such as those indicated above have not been implemented for Bengali. Because powerful deep learning methods require large-scale annotated datasets, and there is a scarcity of data resources in Bengali language. Bengali, on the other hand, has a high linguistic complexity due to its writing system, inflectional morphology, gemination, and a large number of diphthongs and triphthongs.

Only a few research has been conducted on the Bengali language, and SLR53 from OpenSLR is the dataset that is most frequently used. A new Bengali dataset entitled ``Bengali Common Voice Corpus"\cite{alam2022bengali} is published in this ``BUET CSE Fest DL Sprint" competition which contains 231,120 samples. In this study, the test dataset is assessed using a fine-tuned pretrained model with a 5-gram language model. In order to remove unnecessary predictions, some post processing is also used. Even though the outcomes are not very impressive, this work will advance ASR research on Bengali language.

\section{Dataset and Evaluation Metric}

\subsection{Dataset: Bengali Common Voice Corpus v9.0}
\begin{figure*}[t]
\includegraphics[height=2.1in, width=6.8in]{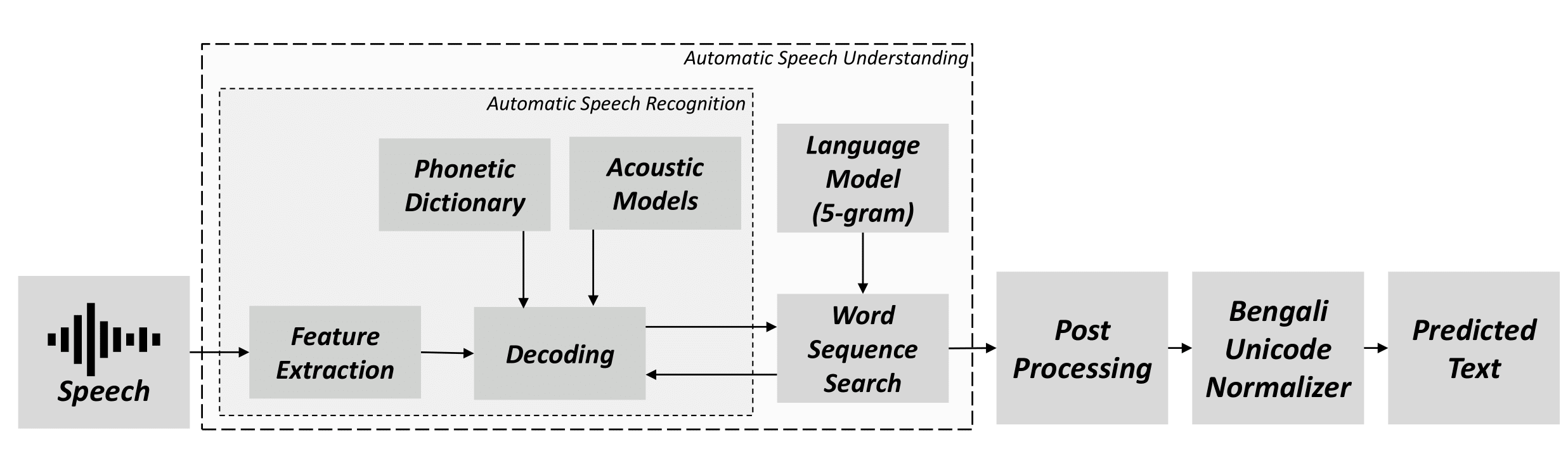}
\centering
\caption{Bengali Automatic Speech Recognition System} 
\label{figure1}
\end{figure*}
The corpus has 231,120 samples from 19,817 contributors resulting in 399 hours (56 hours or 14\% validated by one or more users) of speech recordings. Each audio clip is accompanied with a ‘sentence’ annotation and additional metadata, namely: ‘up votes’, ‘down votes’, ‘age’, ‘gender’, and ‘accents’. The sentence annotation is present for all the audio clips whereas, rest of the metadata comes from only the data contributors that have opted to provide it by logging in to the Mozilla common voice platform. The text corpus contains 135,625 unique sentences. Each sentence has 7.12 words on average and each word contains 3.24 graphemes and 4.95 Unicode characters on average. In this project, the train dataset contains 206950 unique samples, and both the validation and the test dataset contains 7747 unique samples. The samples of these dataset are initially in mp3 format. In this project, all of these samples are converted to wav format for making the process faster. During formatting, the sampling rate of the speech samples was set to 16000 Hz.

\subsection{Evaluation Metric: Levenshtein Mean Distance}
Predictions in the test dataset are evaluated by Levenshtein Mean Distance\cite{heeringa2004measuring} between the predicted sentence and the ground truth sentence. The Levenshtein distance between two sentences is the minimum number of single-character edits (insertions, deletions, or substitutions) required to change one sentence into the other.

The Levenshtein Distance between two strings a, b having length $|a|$ and $|b|$  correspondingly is given by lev(a, b),
\begin{figure}[ht]
\includegraphics[height=1.0in, width=3.4in]{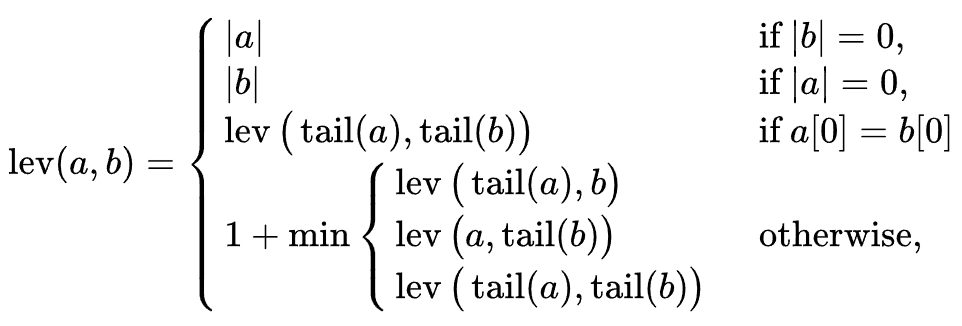}
\centering
\end{figure}

Now, Levenshetin Mean for n samples, where gt[i] and pred[i] indicates ground truth and prediction of sample i, is calculated by,
\begin{figure}[ht]
\includegraphics[height=0.4in, width=3.4in]{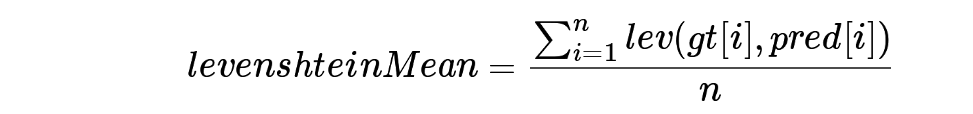}
\centering
\end{figure}

\section{Methodology}

A multilingual speech model called IndicWav2Vec\cite{javed2022towards}, developed by AI4Bharat, is pretrained on 40 Indian languages. In the set of multilingual speech models, this model represents the widest variety of Indian languages. Additionally, it has been fine-tuned for downstream ASR on 9 languages, including Bengali, and achieves cutting-edge performance on 3 open benchmarks, namely MUCS, MSR, and OpenSLR. The pretrained model utilized in this project, "ai4bharat/indicwav2vec v1 bengali," is a fine-tuned version of "facebook/wav2vec2-large-960h-lv60-self" and it was specifically trained on the OpenSLR Bengali dataset. In this project, this model is further fine-tuned on Bengali Common Voice Corpus dataset with only the first 1000 training samples. The feature extractor is kept frozen throughout training. The learning rate was set to 0.0003 and trained for 30 epochs.

The test dataset of 7747 samples is evaluated using this newly fine-tuned version of the model. The KenLM\cite{heafield2011kenlm} 5-gram language model is also utilized in the evaluation section. The library KenLM, developed by Kenneth Heafield, includes two data structures for efficient language model queries that use less memory and time. Language models\cite{buck2014n}, or LMs, are models that assign probability to word sequences. An n-gram is a sequence of n words: a 2-gram (bigram) is a two-word sequence of words like “please turn”, “turn your”, or ’’your homework”, and a 3-gram (trigram) is a three-word sequence of words like “please turn your”, or “turn your homework”. Some unexpected characters are predicted in the evaluation phase due to variances between the train dataset's vocabulary and the pretrained model's vocabulary. To reduce those errors, a post processing technique is applied. A Unicode normalizer is then applied to further improve the assessment performance. Figure \ref{figure1} illustrates the working principle of this system.

\section{Results \& Discussions}
\begin{figure*}[t]
\includegraphics[height=3.4in, width=6.8in]{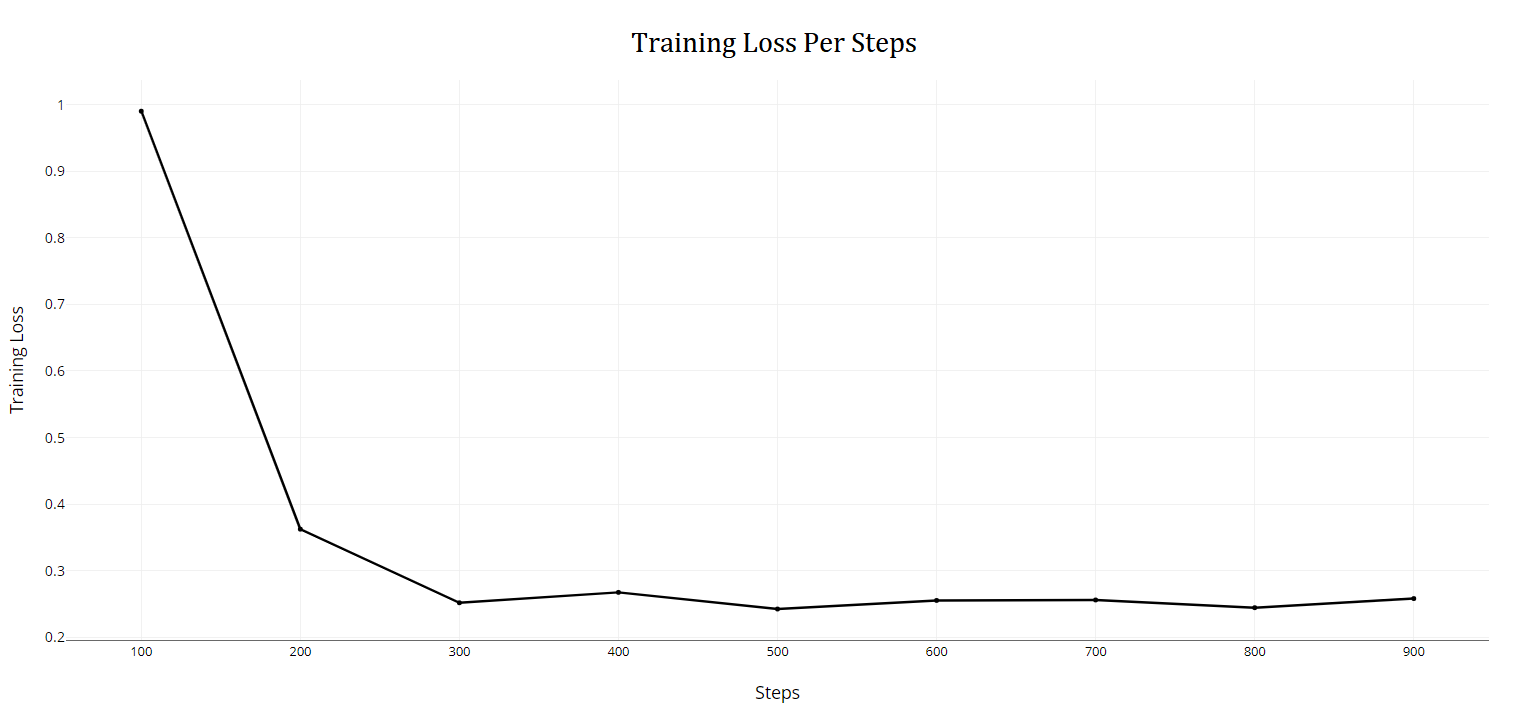}
\centering
\caption{Training Loss Per Steps} 
\label{figure2}
\end{figure*}
First, the test dataset is evaluated using the ``ai4bharat/indicwav2vec-v1-bengali," pretrained model without any training on the train dataset, which performed well with a Levenshtein Mean Distance of 4.175. A Bengali Unicode normalizer and some post-processing techniques were utilized at the time. The first 1000 training examples are used to train this pre-trained model, and 200 validation samples are used to assess performance. Figure \ref{figure2} illustrates the training loss per steps. The loss metric that used is Word Error Rate (WER). The 5-gram language model is employed in the evaluation phase. However, some unnecessary characters are predicted in the evaluation phase, and these are substituted by spaces in the post processing part. Additionally, the last character in the majority of the true samples is ``\textbar", thus a``\textbar" is added to the end of the predicted text. After applying several post-processing methods and a Bengali Unicode normalizer, the test dataset's Levenshtein Mean Distance was found to be 3.819.

The performance of this model can be enhanced in numerous ways. The pretrained model should be trained on the entire train dataset. Because large datasets often help deep learning models train more effectively. Despite the training dataset having more than 207k samples, this model was only trained on 1000 training samples. Therefore, the performance that this model showed is as expected. Special characters like the comma (,), exclamation (!), question mark (?) and others are not taken into account during training, which also had an impact on the model's performance. Besides, Bengali language is very complex in the linguistic perspective due to its writing system, inflectional morphology, gemination, and many diphthongs and triphthongs. This model struggles to make accurate gemination predictions sometimes. Furthermore, most of the time it failed to predict the bengali ``Chandrabindu" character accurately. The Table \ref{wrong_predictions} shows a few typical mistakes.
\begin{table}[ht]
\caption{Some examples of wrong predictions} 
\includegraphics[height=2.0in, width=3.4in]{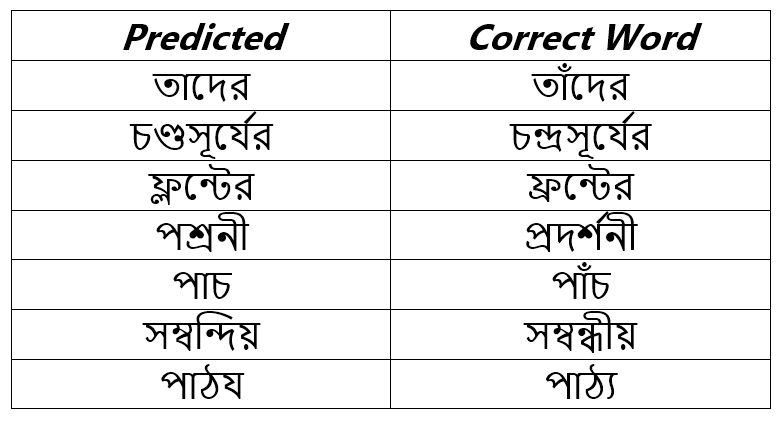}
\centering
\label{wrong_predictions}
\end{table}

Furthermore, most of the dataset samples were not collected in a controlled environment. Many of them are so noisy that even native Bengali speakers could not recognize them. Also, the samples are taken from Bengali people of all around the world. And people of different areas pronounce the same word quite differently. To mitigate this issue, the model should train on large diverse dataset. Therefore, the trained model will perform significantly better if the aforementioned aspects are taken into account during training.

This proposed model was further tested on a new hidden test dataset in the DL Sprint Competition's final round. There were about 4.9k samples in the hidden dataset. This model achieved a Levenshtein Mean Distance of 7.727 on this hidden test dataset.

\section{Conclusions}

In this paper, an ASR system is proposed, which takes a Bengali audio speech file as input and outputs the predicted Bengali sentence. The backbone of the ASR system is the trained model with the 5-gram language model. The output result is improved significantly when the predicted text has been post-processed and normalized using the Bengali Unicode normalizer. On the test dataset, which has 7747 samples, this system is then evaluated and obtained a score of 3.819 in respect of Levenshtein Mean Distance. Only 1000 training samples were used to train the proposed framework due to a GPU runtime issue in the Kaggle environment. The performance may have improved more if the model could have been trained on more effective training samples. This paper demonstrates that ASR in Bengali is feasible, but more research is required to enhance the system's functionality. Additionally, more broad and reliable Bengali speech datasets are required to build a powerful Bengali ASR system. Then, researchers will be very interested in working on Bengali language in the area of artificial intelligence.

\section*{Acknowledgment}

I want to thank the Computer Science \& Engineering Department of Bangladesh University of Engineering \& Technology for hosting this wonderful competition. Additionally, I want to thank Bengali.AI for their tremendous assistance during this competition and for taking the initiative to generate a large dataset of Bengali speech.

\section*{Information Sharing Statement}
The training notebook, inference notebook and the datasets that used in this project are available at \url{https://github.com/tushartalukder/Bengali-ASR-model-using-Wav2Vec2.git}

\bibliographystyle{IEEEtran}
\bibliography{biblio}

\end{document}